\documentstyle[12pt]{article}
\begin{document}
\title{The radial supersymmetry of the  (d+1)-dimensional relativistic 
       rotating oscillators}
\author{Ion I. Cot\u aescu and Ion I. Cot\u aescu Jr.\\        
{\small \it The West University of Timi\c soara,}\\
{\small \it V. P\^ arvan Ave. 4, RO-1900 Timi\c soara, Romania}}

\date{\today}

\maketitle

\begin{abstract}
We study the supersymmetry of the radial problems of the models of quantum 
relativistic rotating oscillators in arbitrary dimensions, defined as 
Klein-Gordon fields in backgrounds with deformed anti-de Sitter metrics. 
It is pointed out that the shape invariance of the supersymmetric partner 
radial potentials leads to simple operators forms of the Rodrigues formulas 
for the normalized radial wave functions.

Pacs 04.62.+v
\end{abstract}
\

\newpage

\section{Introduction}
\

The geometric quantum models in general relativity are devoted to the study of
quantum modes of the free fields in given backgrounds. Of a special interest 
are the analytical solvable models since these can be analyzed in all details 
giving thus information about the relation between the geometry of the 
background and the properties of the specific quantum objects. In this way we  
can better understand how the geometry determines the mean features of the 
whole algebras of obsevables that include the dynamical algebras and the 
operators involved in supersymmetry.  

One of the simplest geometric models is that of the relativistic quantum 
oscillator simulated by the scalar (Klein-Gordon) field in the anti-de Sitter 
background \cite{AIS}. We generalized this model to families of 
(1+1)-dimensional relativistic oscillators \cite{C1} or relativistic rotating 
oscillators in (3+1) \cite{C2} or arbitrary dimensions \cite{C3,C4}. In general, 
the backgrounds of these models have local charts with deformed anti-de Sitter  
or de Sitter metrics. In the case of the deformed anti-de Sitter metrics the 
Klein-Gordon equation lead to standard P\" oschl-Teller problems \cite{PT} in 
(1+1) dimensions \cite{CS} or to similar radial problems in higher dimensions 
\cite{C3,C4}, all of them having countable discrete energy spectra  and 
square integrable energy eigenfunctions. This is the motive why the properties 
of the dynamical algebras as well as the supersymmetry of these models can be 
easily studied. Thus we have shown that all the (1+1) P\" osch-Teller models 
have the same dynamical algebra, $so(1,2)$ \cite{CA}, and good properties of 
supersymmetry and shape invariance which allowed us to write down the Rodrigues 
formulas of the normalized energy eigenfunctions in operator closed form 
\cite{CS}.     

In the present article we should like to continue this study with the 
supersymmetry of the radial problems of the P\" oschl-Teller-type models in 
arbitrary (d+1) dimensions we have recently solved by using traditional methods 
\cite{C4}. Our main objective is to introduce the operators involved in 
supersymmetry and to derive the operator version of the  Rodrigues formulas of 
the normalized radial functions which enter in the structure of the energy 
eigenfunctions. 

In Sec.2 we briefly present the radial problems of the models of rotating 
oscillators in (d+1) dimensions showing that these are P\" oschl-Teller-type 
problems with a very simple parameterization. The supersymmetry of these 
problems and the shape invariance of the supersymmetric partner (superpartner) 
radial potentials is discussed in the next section where we obtain the operator 
form of the mentioned Rodrigues formulas. Sec.4 is devoted to a special  
model of this family which is very similar to the usual nonrelativistic 
harmonic oscillator. The usual expressions of the Rodrigues formulas for the 
radial functions of all these models are given in Appendix. We work in natural 
units with $\hbar=c=1$.

\section{Radial P\" oschl-Teller problems}
\

The  (d+1)-dimensional rotating oscillators are simple geometric models of 
scalar particles freely moving on backgrounds able to simulate oscillatory 
geodesic motions. These  background are static and spherically symmetric 
(central) \cite{C1}  having  static charts with generalized spherical 
coordinates, $r, \theta_{1},..., \theta_{d-1}$,  commonly related with the 
Cartesian ones ${\bf x}\equiv (x^{1}, x^{2},...,x^{d})$ \cite{T}. It is 
convenient to chose the radial coordinate such that $g_{rr}=-g_{00}$ since then 
the radial scalar product is 
simpler \cite{C4}. The metrics of our models \cite{C4} are  one-parameter 
deformations of the AdS metric given by the line elements 
\begin{equation}
ds^{2}=\left( 1+\frac{1}{\epsilon^2}\tan^{2}\hat\omega r\right) (dt^{2}-dr^{2}) 
-\frac{1}{\hat\omega^2}\tan^{2}\hat\omega r\, d\theta^{2}
\end{equation}
where we denote $\hat\omega=\epsilon\,\omega$, $\epsilon\in [0, \infty)$, and  
\begin{equation}
d\theta^{2}={d\theta_{1}}^{2}+\sin^{2}\theta_{1}{d\theta_{2}}^{2}\, ...\, 
+\sin^{2}\theta_{1}\,\sin^{2}\theta_{2}...
\sin^{2}\theta_{d-2}{d\theta_{d-1}}^{2} 
\end{equation}
is the usual line element on the sphere $S^{d-1}$. The deformation parameter 
$\epsilon$ determines the geometry of the background while $\omega$ remains 
fixed. It is clear  that for $\epsilon=1$ we obtain just the AdS metric (with 
the hyperboloid radius $R=1/\omega$) \cite{C3,C4}. An interesting case is 
the model with  $\epsilon= 0$ called  normal oscillator \cite{C4} since 
its line element, 
\begin{equation}
ds^{2}=(1+\omega^{2} r^{2}) (dt^{2}-dr^{2}) -
 r^{2}\, d\theta^{2}\,,
\end{equation}
defines a background where the relativistic quantum motion is similar to that 
of the nonrelativistic harmonic oscillator.  In general, the radial 
domain of any RO is  $D_{r}=[0,\, \pi/2\hat\omega)$ which means 
that the whole space domain is $D=D_{r}\times S^{d-1}$. 
For the models with $\epsilon\not=0$ the time  might satisfy the condition  
$t\in [-\pi/\hat\omega,\pi/\hat\omega)$ as in the AdS case but here we  
consider that $t\in (-\infty,\infty)$ which corresponds to the universal 
covering spacetimes of the hyperbolic original ones.

In these models the oscillating test particle is described by a scalar quantum 
field $\phi$ of mass $M$, minimally coupled with the gravitational 
field. Its quantum modes are given by the particular solutions of the 
Klein-Gordon equation 
\begin{equation}\label{(kg)}
\frac{1}{\sqrt{g}}\partial_{\mu}\left(\sqrt{g}\,g^{\mu\nu}\partial_{\nu}\phi
\right) + M^{2}\phi=0\,, \quad
g=|\det(g_{\mu\nu})|\,, 
\end{equation} 
which, in the case of our models, must be square integrable functions  
\cite{C4} on the domain $D$, orhonormalized  with respect to the relativistic 
scalar product \cite{BD} 
\begin{equation}\label{sc}
\left<\phi,\phi'\right>=i\int_{D}d^{d}x\,\sqrt{g}\,g^{00}\,
{\phi}^{*}\stackrel{\leftrightarrow}{\partial_{0}} \phi'\,.
\end{equation}
The spherical variables  can be separated by using  
generalized spherical harmonics, $Y^{d-1}_{l\,(\lambda)}({\bf x}/r)$. 
These are normalized eigenfunctions of the angular Laplace operator \cite{T},  
\begin{equation}
-\Delta_{S}
Y^{d-1}_{l\,(\lambda)}({\bf x}/r) 
=l(l+d-2)\,Y^{d-1}_{l\,(\lambda)}({\bf x}/r)\,, 
\end{equation} 
corresponding to eigenvalues depending on the {\em angular} quantum number 
$l$ which  takes  only integer values, $0,1,2,....$, selected by the 
boundary conditions on the sphere $S^{d-1}$ \cite{T}. The notation $(\lambda)$ 
stands for a collection of quantum numbers giving the multiplicity of these 
eigenvalues \cite{T}, 
\begin{equation}\label{(gamal)}
\gamma_{l}= (2l+d-2)\frac{(l+d-3)!}{l!\,(d-2)!}\,.
\end{equation}
In general,  the particular solutions of the Klein-Gordon equation of energy 
$E$ (and positive frequency) have the form \cite{C4}  
\begin{equation}\label{(udex)}
\phi^{(+)}_{E,l(\lambda)}(t,{\bf x})=\frac{1}
{\sqrt{2E}} 
(\hat\omega\cot\hat\omega r )^{\frac{d-1}{2}}\, 
R(r)\,Y^{d-1}_{l\,(\lambda)}({\bf x}/r)\, e^{-iEt}\,, 
\end{equation}\label{scr}
involving the radial wave functions $R(r)$ defined  such  that 
the scalar product (\ref{sc}) leads to the simplest {\em radial} scalar product, 
\begin{equation}\label{scrad}
\left<R,R'\right>=\int_{D_{r}}dr\, R^{*}(r)R'(r)\,,
\end{equation}
when the spherical harmonics are normalized to unity.

In Ref.\cite{C4} we have shown that, after the separation of spherical 
variables, the remaining radial Klein-Gordon equation can be treated as an 
independent P\" oschl-Teller problem, with discrete energy spectra and
a very simple parameterization given by the quantum number $l$ and the 
specific parameter 
\begin{equation}
k=\sqrt{\frac{M^{2}}{\hat\omega^{2}\epsilon^2}+\frac{d^2}{4}}+ \frac{d}{2}  
\end{equation}
which concentrates all the other ones, playing thus the role of the main 
parameter of our models. Therefore,  we can consider that the geometric 
parameters ($\omega$ and $\epsilon$) are fixed while the mass of the 
test particle is given by ${M_{k}}^{2}=\epsilon^2\hat\omega^2\,k(k-d)$. For 
this reason we denote the model by $[k]$ understanding that it generates the 
radial problems $(k,l)$, each one having its own sequence of radial 
wave functions,  $R_{k,l,n_{r}}(r)$, labeled by the radial quantum number 
$n_{r}=0,1,2,...$  \cite{C4}.  With these notations, the radial equation 
for a given pair $(k,l)$ can be put in the form   
\begin{equation}\label{(radeq)}
\left[-\frac{1}{\hat\omega^2}\frac{d^2}{dr^2}+\frac{2s(2s-1)}{\sin^{2}
\hat\omega r}+
\frac{2p(2p-1)}{\cos^{2}\hat\omega r}\right]R_{k,l,n_{r}}=\nu^{2} 
R_{k,l,n_{r}}
\end{equation}
where the values of the parameters $s$ and $p$ corresponding to the regular 
modes are 
\begin{equation}\label{(eqps)}
2s=l+\delta\,, \quad 
2p=k-\delta\,,\quad \delta=\frac{d-1}{2}
\end{equation}
while the quantization condition reads
\begin{equation}
\nu=2(n_{r}+s+p)\,.
\end{equation}
Hereby it results that the discrete energy levels, 
\begin{equation}\label{(e)}
E_{k,n,l}^{2}=\hat\omega^{2}(k+n)^{2}+\hat\omega^{2}(\epsilon^{2}-1)
\left[k(k-d)-\frac{1}{\epsilon^2} l(l+d-2)\right].
\end{equation}
depend on the  main quantum number, $n=2n_{r}+l$.  
If $n$ is even then $l=0,2,4,...,n$ while for odd $n$ we have $l=1,3,5,...,n$. 
In both cases the degree of degeneracy of the level $E_{k,n,l}$ is given by 
(\ref{(gamal)}). 

In the following we use only the quantum number $n_{r}$  since this labels the 
sequences of radial functions $R_{k,l.n_{r}}$ which form the {\em energy bases} 
of the  P\" oschl-Teller radial problems $(k,l)$ in the Hilbert space 
${\cal H}_{r}$, of the square integrable functions with respect to the radial 
scalar product (\ref{scrad}). We specify that all our radial functions 
accomplish the boundary conditions $R(0)=R(\pi/2\hat\omega)=0$ such that the 
Hermitian conjugation of the  operators on ${\cal H}_{r}$  can be correctly 
defined with respect to this scalar product 
(e.g., ${\partial_{r}}^{\dagger}=-\partial_{r}$). Thus we obtain a familiar 
approach similar to that of the non-relativistic radial problems in the 
coordinate representation of the Schr\" odinger picture.

\section{Radial supersymmetry}
\

The supersymmetric formalism of the radial problems $(k,l)$ can be constructed 
like that of the relativistic (1+1)-dimensional P\" oschl-Teller models 
\cite{CS}. To this end it is convenient to introduce the operator   
\begin{equation}
\{{\bf \Delta}[V]R\}(r)=\left(-\frac{d^2}{dr^2}+V(r)\right)R(r),
\end{equation}
which should play the same role as the Hamiltonian of the one-dimensional 
nonrelativistic problems.  

Starting with  the normalized ground-state radial function 
\begin{equation}\label{(eg)}
R_{k,l,0}(r)=\sqrt{2\hat\omega}\left[
\frac{\Gamma(k+l+1)}{\Gamma(l+\frac{d}{2})
\Gamma(k+1-\frac{d}{2})}\right]^{\frac{1}{2}}
\sin^{2s}\hat\omega r \cos^{2p}\hat\omega r
\end{equation}
we obtain the radial superpotential
\begin{equation}\label{(sup)}
W(k,l,r)=-\frac{1}{R_{k,l,0}(r)}\frac{dR_{k,l,0}(r)}{dr}=
\hat\omega[2p\tan\hat\omega r-2s\cot \hat\omega r]
\end{equation} 
that help us to find the superpartner radial potentials
\begin{eqnarray}
V_{\pm}(k,l,r)&=&\pm\frac{dW(k,l,r)}{dr}+W(k,l,r)^{2}\nonumber\\
&=&\hat\omega^{2}\left[\frac{2s(2s\pm 1)}{\sin^{2}\hat\omega r}
+\frac{2p(2p\pm 1)}{\cos^{2}\hat\omega r}-(2s+2p)^2\right]\,.
\end{eqnarray}
Now  Eq.(\ref{(radeq)}) can be rewritten as 
\begin{equation}\label{(o1)}
{\bf \Delta}[V_{-}(k,l)]R_{k,l,n_{r}}=d_{k,l,n_{r}} R_{k,l,n_{r}}
\end{equation}
where 
\begin{equation}\label{(dif)}
d_{k,l,n_{r}}=\hat\omega^{2}[{\nu}^{2}-(\nu_{|n_{r}=0})^{2}]=
4 \hat\omega^2 n_{r}(n_{r}+k+l)
\end{equation}
satisfies $d_{k,l,0}=0$.

Furthermore, according to the standard procedure \cite{S}, we  define the 
pair of adjoint operators, ${\bf A}_{k,l}$ and ${\bf A}_{k,l}^{\dagger}$, 
having the action 
\begin{eqnarray}
({\bf A}_{k,l}R)(r)&=&\left(\frac{d}{dr}
+W(k,l,r)\right)R(r),\label{(aunu)}\\ 
({\bf A}_{k,l}^{\dagger}R)(r)&=&\left(-\frac{d}{dr}+
W(k,l,r)\right)R(r)
\end{eqnarray}
and allowing us to write       
\begin{equation}
{\bf \Delta}[V_{-}(k,l)]={\bf A}_{k,l}^{\dagger}{\bf A}_{k,l} , \qquad
{\bf \Delta}[V_{+}(k,l)]={\bf A}_{k,l}{\bf A}_{k,l}^{\dagger}.
\end{equation}

Let us observe now that the radial potentials $V_{-}(k,l)$ and $V_{+}(k,l)$ are 
shape invariant since
\begin{equation}\label{(sh)}
V_{+}(k,l,r)=V_{-}(k+1,l+1,r)+4\hat\omega^{2}(k+l+1).
\end{equation}
Consequently, we can verify that 
\begin{equation}\label{(o2)}
{\bf \Delta}[V_{+}(k,l)]R_{k+1,l+1,n_{r}-1}=d_{k,l,n_{r}}R_{k+1,l+1,n_{r}-1},
\quad n_{r}=1,2,...\quad,
\end{equation}
which means that the spectrum of the operator ${\bf \Delta}[V_{+}(k,l)]$ 
coincides with that of ${\bf \Delta}[V_{-}(k,l)]$, apart from the lowest 
eigenvalue  $d_{k,l,0}=0$. From (\ref{(o2)}) combined with  (\ref{(o1)}) it 
results that the normalized radial functions satisfy
\begin{eqnarray}
{\bf A}_{k,l}R_{k,l,n_{r}}&=&\sqrt{d_{k,l,n_{r}}}R_{k+1,l+1,n_{r}-1}\,,
\label{(aa1)}\\
{\bf A}_{k,l}^{+}R_{k+1,l+1,n_{r}-1}&=&\sqrt{d_{k,l,n_{r}}}R_{k,l,n_{r}}\,.
\label{(aa2)} 
\end{eqnarray}
Hence, we have obtained the desired relation between the energy bases of the 
radial problems  $(k,l)$ and $(k+1,l+1)$  which have the superpartner radial 
potentials $V_{-}(k,l)$ and $V_{+}(k,l)$ respectively. In general, any 
normalized radial function of the basis $(k,l)$ of the model $[k]$ can be 
written as
\begin{eqnarray}\label{Rod}
R_{k,l,n_{r}}&=&\frac{1}{(2\hat\omega)^{n_{r}}}\left[\frac{\Gamma(n_{r}+k+l)}
{n_{r}!\,\Gamma(2n_{r}+
k+l)}\right]^{\frac{1}{2}}{\bf A}_{k,l}^{\dagger}{\bf A}_{k+1,l+1}^{\dagger}
\cdots\nonumber\\
&&\cdots {\bf A}_{k+n_{r}-1,l+n_{r}-1}^{\dagger}R_{k+n_{r},l+n_{r},0}
\label{(Rod)}.
\end{eqnarray}
where $R_{k+n_{r},l+n_{r},0}$ is the normalized ground-state radial function of 
the energy basis $(k+n_{r},l+n_{r})$ of the model $[k+n_{r}]$, given by 
Eq.(\ref{(eg)}). Thus we  obtain the operator form of the Rodrigez formula of 
the radial functions as in the case of the (1+1)-dimensional relativistic 
P\" oschl-Teller models \cite{CS}. 

\section{The normal oscillator}
\

A special case is that of $\epsilon \to0$ when one obtains the 
normal oscillator which has similar radial  functions as those of the 
nonrelativistic harmonic oscillator \cite{C4}.  
In this limit we have $\hat\omega\to 0$, $k\to \infty$, but 
$\epsilon^{2}k\to M/\omega$, such that the superpotential becomes
\begin{equation}
W(l,r)=\lim_{\epsilon\to 0}W(k,l,r)=  
\omega M r-\frac{l+\delta}{r}  
\end{equation}
giving the superpartner radial potentials 
\begin{eqnarray}\label{vecuc}
V_{\pm}(l,r)&=&\lim_{\epsilon\to 0}V_{\pm}(k,l,r)\nonumber\\
&=& \omega^2 M^2 r^2+ \frac{(l+\delta)(l+\delta\pm 1)}{r^2}-
\omega M[2(l+\delta)\mp 1]
\end{eqnarray}
which are shape invariant since
\begin{equation}\label{si}
V_{+}(l,r)=V_{-}(l+1,r)+4\omega M\,.
\end {equation}
Thus we see that the shape of the superpartner potentials is given  only 
by the parameter $l$. Therefore, we can consider all the other parameters 
(including the mass) as being fixed and denote the radial problems of the 
normal oscillator by $(l)$. 

Since $\lim_{\epsilon\to 0}d_{k,l,n_{r}}=4\omega M n_{r}$ 
it is convenient to  introduce the new operators 
\begin{equation}
{\bf a}_{l}=\frac{1}{2\sqrt{\omega M}}\lim_{\epsilon\to 0}{\bf A}_{k,l}
\end{equation}
which have the action
\begin{eqnarray}
({\bf a}_{l}R)(r)&=&\frac{1}{2\sqrt{\omega M}}\left(\frac{d}{dr}+\omega M r-
\frac{l+\delta}{r}\right)\,,\\ 
({\bf a}^{\dagger}_{l}R)(r)&=&\frac{1}{2\sqrt{\omega M}}\left(-\frac{d}{dr}+
\omega M r-\frac{l+\delta}{r}\right)\,, 
\end{eqnarray}
and satisfy
\begin{equation}
\Delta[V_{-}]=4\omega M\,{{\bf a}_{l}}^{\dagger} {\bf a}_{l}\,,\quad
\Delta[V_{+}]=4\omega M\,{\bf a}_{l}{{\bf a}_{l}}^{\dagger}\,.
\end{equation}
As in the previous cases, from Eqs.(\ref{vecuc}) and (\ref{si}) we deduce that 
these operators have the same spectrum (apart from the lowest eigenvalue) and, 
consequently, we can write      
\begin{equation}
{\bf a}_{l}R_{l,n_{r}}=\sqrt{n_{r}}R_{l+1,n_{r}-1}\,,\quad
{\bf a}^{\dagger}_{l}R_{l+1,n_{r}-1}=\sqrt{n_{r}}R_{l,n_{r}}\,.
\end{equation}
Finally we obtain the operator form of the Rodrigues formula,
\begin{equation}\label{brr}
R_{l,n_{r}}=\frac{1}{\sqrt{n_{r}!}}\,
{\bf a}^{\dagger}_{l}\,{\bf a}^{\dagger}_{l+1}\cdots
{\bf a}^{\dagger}_{l+n_{r}-1}\,R_{l+n_{r},0}
\end{equation}
which express any radial function of the problem $(l)$ in terms of the 
ground-state radial function of the problem $(l+n_{r})$. Hereby it result the 
usual Rodrigues formula of the radial functions of the normal oscillator (as 
given in Appendix).

\section{Concluding remarks}
\

The radial problems of our rotating oscillators have two interesting properties. 
The first one refers to the relation among symmetries and supersymmetries. 
Thus we have seen that all of our models, apart from that with $\epsilon=0$, 
have the same properties concerning the supersymmetry and  shape invariance of 
the radial potentials, despite of the fact that their backgrounds have 
different spacetimes symmetries. Indeed, all the models with $\epsilon 
\not=1$ have central backgrounds with the symmetry given by the group 
$T(1)\otimes SO(d)$ of time translations and space rotations. On the other 
hand, the background of the model with $\epsilon=1$ is just the 
(d+1) dimensional anti-de Sitter spacetime which is the homogeneous space 
of the group $SO(d,2)$ that coincides with the isometry group of  this 
background. Thus we see that problems with different spacetime symmetries have 
similar behaviors from the point of wiev of the supersymmetry and shape 
invariance.

The second property which is worth pointing out concerns the parameterization 
of these models. The radial problems studied here depend on two parameters of 
different nature, $l$ which is the quantum number of the whole angular momentum 
and the usual parameter $k$. What is interesting is that both these 
parameters have the same behavior from the point of wiev of the shape 
invariance. More precisely the problems with superpartner radial potentials 
have simultaneously $\Delta l=\pm 1$ and $\Delta k=\pm 1$ even though these 
parameters are of different origins.       
Thus we find similar properties with those studied in the simplest case of 
the (1+1) models of relativistic oscillators \cite{CS}.   

\appendix

\section{Rodrigues formulas}
\

Starting with the observation that, according 
to Eqs.(\ref{(aunu)}) and (\ref{(sup)}), we have
\begin{equation}
({\bf A}_{k,l}R)(r)= (\sin\hat\omega r)^{2s}(\cos\hat\omega r)^{2p}\frac{d}{dr}
 (\sin\hat\omega r)^{-2s}(\cos\hat\omega r)^{-2p}\, R(r)\,,
\end{equation}
it is not difficult to show that Eq.(\ref{Rod}) gives the Rodrigues formula 
of the normalized radial functions in the new variable $u=\cos2\hat\omega r$. 
This is    
\begin{eqnarray}
&&R_{k,l,n_{r}}(u)=\frac{(-1)^{n_{r}}}{2^{\frac{k+l}{2}+n_{r}}}
\left[\frac{2\hat\omega\, (2n_{r}+k+l)\,\Gamma(n_{r}+k+l)}
{n_{r}!\,\Gamma(n_{r}+l+\frac{d}{2})\,\Gamma(n_{r}+k+1-\frac{d}{2})}
\right]^{\frac{1}{2}}\nonumber\\
&&\times (1-u)^{-\frac{l+\delta-1}{2}}(1+u)^{-\frac{k-\delta-1}{2}}
\frac{d^{n_{r}}}{du^{n_{r}}} (1-u)^{l+\delta+n_{r}-\frac{1}{2}}
(1+u)^{k-\delta+n_{r}-\frac{1}{2}}\,.
\end{eqnarray}

In the case of the normal oscillator we have 
\begin{equation}
({\bf a}_{l}R)(r)=\frac{1}{2\sqrt{\omega M}}r^{l+\delta}e^{\omega Mr^2/2}
\frac{d}{dr} 
r^{-(l+\delta)}e^{-\omega Mr^2/2}R(r)
\end{equation}
and
\begin{equation}
R_{l,0}(r)=\lim_{\epsilon\to 0}R_{k,l,0}(r)=
\left[\frac{2(\omega M)^{l+\frac{d}{2}}}{\Gamma(l+\frac{\delta}{2})}
\right]^{\frac{1}{2}} r^{l+\delta}e^{-\omega M\,r^2/2}
\end{equation}
so that Eq.(\ref{brr}) gives the usual Rodrigues formula of the normalized 
radial functions of the normal oscillator,
\begin{equation}
R_{l,n_{r}}(z)=\frac{(4\omega M)^{\frac{1}{4}}}{\sqrt{n_{r}!\,
\Gamma(n_{r}+l+\frac{d}{2})}}\,
z^{-\frac{l+\delta-1}{2}}e^{z/2}\frac{d^{n_{r}}}{dz^{n_{r}}} 
z^{n_{r}+l+\delta-\frac{1}{2}} e^{-z}\,,
\end{equation}
where $z=\omega M r^2$. 

Thus the supersymmetry and shape invariance help us to easily recover our 
previous results obtained by using traditional methods \cite{C4}.


\end{document}